\documentclass[12pt,preprint]{aastex}

\shorttitle{Be Star Disks in Vertical Hydrostatic Equilibrium}
\shortauthors{T.\ A.\ A.\ Sigut, M.\ A.\ McGill \& C.\ E.\ Jones}

\begin{document}

\title{Be Star Disk Models in Consistent Vertical Hydrostatic Equilibrium}

\author{T.\ A.\ A.\ Sigut, M.\ A.\ McGill and C.\ E.\ Jones\\
Department of Physics and Astronomy, The University 
of Western Ontario, London, Ontario, N6A 3K7, Canada 
\email{asigut@astro.uwo.ca} \email{mmcgill@astro.uwo.ca} \email{cjones@astro.uwo.ca}}

\slugcomment{Accepted for Publication in The Astrophysical Journal, \\
May 8, 2009}

\begin{abstract}

A popular model for the circumstellar disks of Be stars is that of a
geometrically thin disk with a density in the equatorial plane that drops
as a power law of distance from the star. It is usually assumed that
the vertical structure of such a disk (in the direction parallel to the
stellar rotation axis) is governed by the hydrostatic equilibrium set by
the vertical component of the star's gravitational acceleration.  Previous
radiative equilibrium models for such disks have usually been computed
assuming a fixed density structure. This introduces an inconsistency as
the gas density is not allowed to respond to temperature changes and the
resultant disk model is not in vertical, hydrostatic equilibrium. In
this work, we modify the {\sc bedisk} code of \citet{sig07} so that
it enforces a hydrostatic equilibrium consistent with the temperature
solution. We compare the disk densities, temperatures, H$\alpha$ line
profiles, and near-IR excesses predicted by such models with those
computed from models with a fixed density structure. We find that the
fixed models can differ substantially from the consistent hydrostatic
models when the disk density is high enough that the circumstellar disk
develops a cool ($T\lesssim10,000\;$K) equatorial region close to the
parent star.  Based on these new hydrostatic disks, we also predict an
approximate relation between the (global) density-averaged disk
temperature and the $T_{\rm eff}$ of the central star, covering the full
range of central Be star spectral types.

\end{abstract}


\section{Introduction}

Be stars are non-supergiant B stars that currently show, or have shown
in the past, emission in one or more of the hydrogen Balmer lines,
typically H$\alpha$ \citep{por03}. Other common characteristics of
Be stars include an infrared excess \citep{wat86} and linear continuum
polarization at the level of approximately one percent \citep{wat92}.
These observations can be explained by the presence of a non-spherical
distribution of circumstellar material surrounding the central B~star
\citep{woo97}. Be
stars have recently been resolved interferometrically in the optical
\citep{qui93,tyc05,tyc06} to conclusively show that this circumstellar
material is in the form of a thin equatorial disk, a form originally
championed by \citet{poe78}.  Typically the spectra of Be stars are
consistent with this thin disk in near Keplerian rotation in which the
equatorial density drops as a power-law in radius.  In such models,
the vertical structure of the gas (perpendicular to the disk) is often
assumed to be in isothermal hydrostatic equilibrium. In this case, the overall
disk density, expressed in cylindrical co-ordinates $(R,Z)$ where
the $Z$ direction is parallel to the rotation axis of the star, is
\begin{equation} 
\label{eq:rho} 
\rho(R,Z) =
  \rho_o \left(\frac{R_*}{R}\right)^{n} e^{-\left(\frac{Z}{H}\right)^2}
               \;.
\end{equation} 
Here $R_*$ is the stellar radius, $\rho_o$ is the density of the inner
edge of the disk in the equatorial plane (often termed the ``base density"
of the disk), and $H$ is a vertical scale-height that depends upon the
distance, $R$, the stellar mass, $M_*$, and a disk temperature,
$T_o$, assumed valid for all $Z$ at that distance.  Such a density model
balances the vertical component of the star's gravitational acceleration
(parallel to the rotation axis) with the pressure set by the temperature
$T_o$. If $T_o$ is a constant for all radial distances, the weakening
of the vertical gravitational acceleration with $R$ causes $H\propto
R^{3/2}$ and the disk flares with increasing distance.  Here $\rho_o$
and the power-law index, $n$, are the parameters that determine the
overall density of the disk (together with the assumed value for $T_o$
which is often simply set to a constant fraction of the $T_{\rm eff}$
of the central star). Typically one fixes these parameters by comparing
to the observational constraints cited above. By matching the infrared
excess of a wide sample of Be stars, \citet{wat87} and \citet{dou94}
find power-law indexes in the range of $n=2.0$ to $5.0$\footnote{Both
of these analysis are based on the disk model of \citet{wat86} which
assumes a density distribution of the form $\rho(r)=\rho_o (R_*/r)^n$
in a disk of finite opening angle. Here $r$ is the distance from the
centre of the star.  This density distribution thus {\it differs\/}
from Eq.~\ref{eq:rho}.}.  Theoretically, \citet{por99} notes that
an isothermal, viscous disk requires $n\geq3.5$ for outflow, and
\citet{jon07} find that a hydrodynamical simulation of outflowing viscous
disks in which the thermal structure of the disk is taken into account
\citep[following][]{jon04} predicts $n$ in the range of $3.0$ to $3.5$
in the inner portion of the disk, $R\lesssim10\,R_*$, where hydrogen
spectrum likely forms.

While empirically probing the temperature structure within Be star
disks is a difficult observational challenge, several theoretical
models of the temperature structure of such disks now exist
\citep{car06,sig07}. Typically, for low densities, the disks are
nearly isothermal. However, as the density rises (i.e.\ as $\rho_o$ is
increased), a cool equatorial region develops close to the star and the
disk gas becomes far from isothermal in the vertical direction, even at a
single $R$. Such models are fundamentally inconsistent because the derived
temperature structure can no longer be consistent with the assumption
of a vertically isothermal gas as required by Eq.~\ref{eq:rho}. As
observational diagnostics can be very sensitive to the gas density in
the disk, such an inconsistency may have observational consequences
in terms of the predicted emission line profiles and strengths, the
predicted IR excess, and the predicted linear polarization signature.
For example, \citet{car06} suggest that the fundamental limitation to
\citet{wat86} technique for determining disk density from the slope of
the IR continuum is the a priori assumption of disk geometry via the
adoption of a constant disk opening angle.

It is the purpose of the present work to eliminate this inconsistency
between the temperature and density structure of the disk.  Radiative
equilibrium models are constructed in which the vertical disk density
structure is in a hydrostatic equilibrium consistent with the computed
temperature distribution.

\section{Theory}

We assume an axisymmetric disk described by the cylindrical co-ordinates $(R,Z)$.
We shall also assume that the disk density in the equatorial plane ($Z=0$) is
a known function of $R$ of the form
\begin{equation}
\label{eq:PL}
\rho(R,0)=\rho_o\,\left(\frac{R_*}{R}\right)^{n} \;,
\end{equation}
where $R_*$ is the stellar radius;
$\rho_o$ and the power-law index $n$ are free parameters that fix the
density structure of the disk. Thus at any location $R$, the density $\rho(R,0)$
is known, and we shall denote this density simply as $\rho(0)$. If the gas at this
location is in
vertical hydrostatic equilibrium, then the vertical pressure gradient
must satisfy the equation of hydrostatic equilibrium,
\begin{equation}
\label{eq:HE}
\frac{dP}{dz}= - \rho\,g_{\rm z} \;.
\end{equation}
Here $g_{\rm z}$ is the vertical (or z-component) of the star's gravitational acceleration 
at location $R$, namely
\begin{equation}
\label{eq:gz}
g_z = GM_*\,\frac{Z}{(R^2+Z^2)^{3/2}} \;,
\end{equation}
where $M_*$ is the mass of the central star. In all cases, $\rho_o$ in Eq.~\ref{eq:PL}
is so small
that the mass of the disk is completely negligible compared to that of the star.
The pressure can be eliminated from the hydrostatic equation as the gas is assumed
to obey the perfect gas law,
\begin{equation}
P=\frac{\rho}{\mu m_H}\,kT \;.
\end{equation}
This introduces the vertical temperature distribution into the
problem. In this equation, $P$, $\rho$, $T$, and $\mu$ are functions of $Z$
(the mean-molecular weight, $\mu$, depends on $Z$ because it is determined by the
ionization state of the gas). Using the perfect gas law to eliminate
the pressure from Eq.~\ref{eq:HE}, and using Eq.~\ref{eq:gz}, we find that
\begin{equation}
\label{eq:HERHO}
\frac{1}{\rho}\,\frac{d\rho}{dz} = -\alpha(Z)\,\frac{Z}{(R^2+Z^2)^{3/2}} \;,
\end{equation}
where $\alpha(Z)$ is the function
\begin{equation}
\label{eq:alpha}
\alpha(Z) = GM_*\,\frac{\mu(Z) m_{\rm H}}{kT(Z)} \;.
\end{equation}
Using the boundary condition that the density at $Z=0$ is equal to
$\rho(0)$, Eq.~\ref{eq:HERHO} can be numerically integrated assuming
that the functions $T(Z)$ and $\mu(Z)$ are known. Typically a numerical
solution is required as the temperature $T(Z)$ is available only at a fixed
number of vertical grid points via a radiative equilibrium solution. In
the sections to follow, we shall use the {\sc bedisk} code of \citet{sig07}
to compute the thermal structure of several Be disks so that the vertical
hydrostatic equation can be numerically integrated at each distance from
the star.

The analytic solution of Eq.~\ref{eq:HERHO}, seen in Eq.~\ref{eq:rho}, is
obtained when the vertical temperature variation can be replaced by a constant
temperature $T_o$ (with the mean-molecular weight also assumed to be a constant, $\mu_o$).
In this case, Eq.~\ref{eq:alpha} is just the parameter $\alpha_o$, and
Eq.~\ref{eq:HERHO} is readily integrated to give
\begin{equation}
\label{eq:den_exact}
\rho(Z) = \rho(0) \, e^{-\alpha_o\left(\frac{1}{R}-\frac{1}{\sqrt{R^2+Z^2}}\right)} \;.
\end{equation}
In the case of a geometrically thin disk in which $Z/R\ll 1$, applicable to the Be stars,
this result simplifies to
\begin{equation}
\rho(Z) = \rho(0) e^{ -\frac{\alpha_o}{2R}\left( \frac{Z}{H} \right)^{2} } \;.
\end{equation}
Comparing to Eq.~\ref{eq:rho}, the scale-height H is given by
\begin{equation}
\label{eq:scale_height}
H=\sqrt{\frac{2R^3}{\alpha_o}} \;,
\end{equation}
which gives the cited $H\propto R^{3/2}$ disk flaring with distance. Thus
the assumption of vertically isothermal gas in the Be disk is required
to reproduce the density structure of Eq.~\ref{eq:rho}.

\section{Computational Procedure}

To obtain the temperature structure of various Be star disk models,
we use the {\sc bedisk} code of \citet{sig07}.  This code has been
successfully used to interpret a wide range of Be star observational
signatures \citep[for example, see][]{tyc08,jon08,jon09}.  The {\sc
bedisk} code solves the radiative equilibrium problem for the disk gas
by balancing the heating and cooling rates computed for a user-specified
set of atomic models.  Energy input to the disk is assumed to come from
the radiation of the central star.  This photoionizing radiation field
is represented as a sum of the direct component from the central star
and the diffuse component from the disk itself, which is treated by the
on-the-spot (OTS) approximation \citep{ost89}.
%
%
\cite{sig07} present results for a direct (but still approximate)
treatment of the diffuse component to demonstrate its affect on the
thermal structure of the disk. They find that even for a dense disk with
$\rho_o=5\cdot10^{-11}\;\rm g\,cm^{-3}$ and an $R^{-2.5}$ radial drop-off,
the additional heating provided by the diffuse component increases the
temperature in the inner equatorial plane by only 10-20\% and conclude
that the OTS approximation gives good results.

%
%
Limb darkening of the central star is accounted for in the direct
photoionizing radiation field, but the gravitational darkening and
geometric distortion implied by rapid rotation is yet to be implemented.
Heating of the gas is via photoionization and collisional excitation.
Cooling of the gas is via recombination and collisional de-excitation. The
atomic level populations required to compute the heating and cooling
rates were found by solving the requisite statistical equilibrium
equations. The transfer of line radiation is handled through the escape
probability approximation. See \cite{sig07} for further details.

To achieve a solution that is both in radiative equilibrium and in
a vertical hydrostatic equilibrium consistent with the temperature
structure, some modification to the {\sc bedisk} code is required.
The initial density distribution is simply that of Eq.~\ref{eq:rho}.
{\sc bedisk} then obtains the disk temperatures on a fixed grid specified
by the cylindrical coordinates $(R_i,Z_{ij})$, with $i=1,2,\ldots,n_r$
and $j=1,2,\ldots,n_z$ where $R_{i+1}>R_i$ and $Z_{i,j+1}>
Z_{i,j}$. The solution starts closest to the star (at $i=1$) and
statistical and radiative equilibrium is solved for each $Z_{1,j}$ for
$j=n_z,n_z-1,\ldots,1$, i.e.\ by proceeding down towards the equatorial
plane. Then Eq.~\ref{eq:HERHO} is numerically integrated using the density
in the equatorial plane as the boundary condition.  As a result of this
integration, the total gas density for $i=1$ and all $j=1,2,\ldots,n_z$
is updated. As the gas density at each $Z_{1,j}$ has been changed,
radiative equilibrium must be resolved to obtain a new temperature
distribution. This process is iterated until the maximum fractional
change in the gas density drops below 1\%. Once this has happened, the
calculation proceeds to the next radial distance in the disk, $i=2$,
and so on.

If, on average, $N_{\rho}$ density iterations are required to meet
the convergence tolerance, then the entire calculation for consistent
vertical hydrostatic equilibrium takes $N_{\rho}$ times that required for
the radiative equilibrium solution with the fixed density structure of
Eq.~\ref{eq:rho}. As $N_{\rho}\approx 5-10$, a significant lengthening
of the execution time of {\sc bedisk} occurs. Finally, we have found
that the various iterative estimates for the density often tend to oscillate
around the true value, and it is generally advantageous to update the
grid densities with a relaxed estimate that is the average of the
old value and the new predicted estimate.

\section{Results}

To demonstrate typical differences between disks computed with a fixed,
isothermal density structure (referred to as ``fixed" models) and those
computed with consistent vertical hydrostatic equilibrium (referred
to as ``hydrostatic" models), we consider pure hydrogen and helium
envelopes surrounding the three central stars with parameters found in
Table~\ref{tab:stellar_param}. These parameters represent central stars
with spectral types of approximately B0, B2, and B5. While the {\sc
bedisk} code can handle multiple atomic models, and thus can compute
the radiative equilibrium solution for a gas with a solar composition,
such calculations are much more computationally intensive than the
hydrogen/helium models considered here. Nevertheless, the hydrogen/helium
models are completely adequate to illustrate the main differences between
the fixed and hydrostatic models over a wide range of disk parameters. For
the atomic models, we have adopted the 15 level H\,{\sc i} and 13 level
He\,{\sc i} models of \citet{sig07}; H\,{\sc ii} and He\,{\sc ii}/{\sc
iii} were represented as single levels. While the inclusion of helium
has only a small affect on the thermal solution, it allows for a more
realistic mean molecular weight for the gas. The total helium number
density was assumed to be 0.1 of the total hydrogen number density.

For all calculations, the disk was represented by $N_r=84$ and $N_z=50$
grid points with $R_{\rm max}/R_*=50$. The power law density index for the
equatorial density (Eq.~\ref{eq:PL}) was taken to be $n=3.5$ in all cases.

\begin{deluxetable}{lrrr}
\tablewidth{0pt}
\tablecaption{Adopted parameters for the central B stars.\label{tab:stellar_param}}
\tablehead{
\colhead{Parameter} & \colhead{B0}  & \colhead{B2} & \colhead{B5}
}
\startdata
Radius ($\rm R_{\sun}$)         & 10.0          &   7.0    &  5.0   \\
Mass ($\rm M_{\sun}$)           & 17.0          &  12.0    &  9.0   \\
$\log_{10}(L / L_{\sun}$) & 4.5           &   3.8    &  3.0   \\
$T_{\rm eff}\;$(K)                   & 25000         & 20000    & 15000  \\
$\log(g)$                       &  4.0         &   4.0    &  4.0   \\ 
\enddata
\vspace{0.1in}
\end{deluxetable}

In the discussion that follows, we first outline the detailed differences
between the predicted temperature and density distributions in the
disk models. Next we consider how well the fixed models can predict
the density-weighted disk temperature given an appropriate choice for
the $T_o$ parameter. Finally, we examine the differences in selected
observational predictions of the models, namely in the H$\alpha$ line
profile and in the near-infrared excess, both of which have been used
by previous investigators to determine the density structure of Be star
circumstellar disks.

\subsection{Disk Temperatures and Densities}
\label{sec:tandrho}

\begin{figure}
\epsscale{1.0}
\plotone{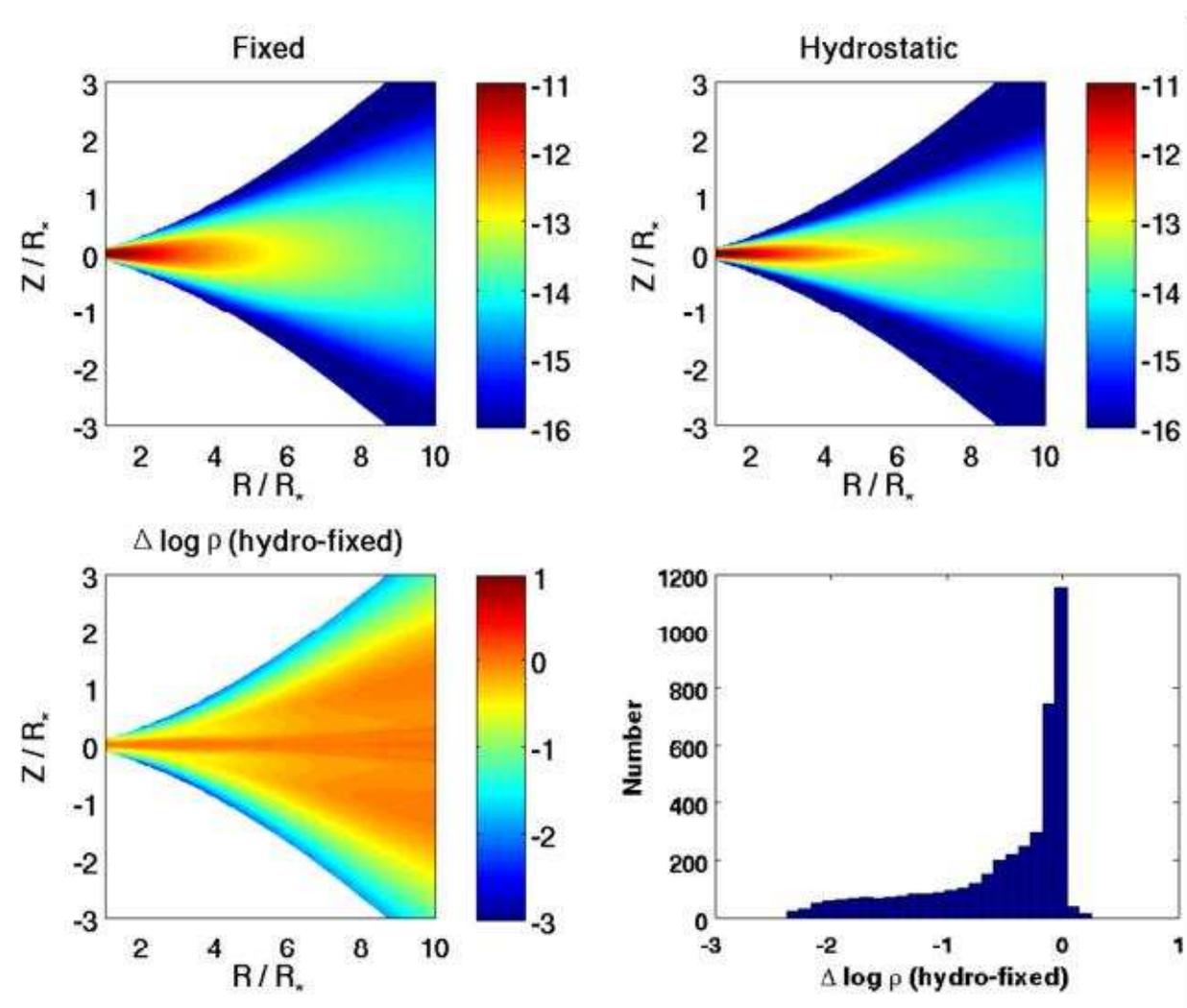}
\caption{The density structure of a fixed (upper
left) and hydrostatic (upper right) disk with $\rho_o=5\cdot10^{-11}\,
\rm g\,cm^{-3}$ and $n=3.5$ for the B0 model of Table~\ref{tab:stellar_param}.
The lower-left panel shows the differences in $\log(\rho)$ directly,
while the lower-right panel gives a histogram
of these differences over all 4200 grid points.\label{fig:egden}}
\vspace{0.1in}
\end{figure}

Figure~\ref{fig:egden} compares the predicted density structure
of fixed and hydrostatic disks computed for the B0 model with
$\rho_o=5\cdot10^{-11}\;\rm g\,cm^{-3}$. In the case of the fixed
model, an isothermal temperature of $T_o=13,500\;$K was chosen for
Eq.~\ref{eq:rho} (the reason for this choice will be discussed below).
The lower-left panel of this figure shows the difference in the logarithm
of the predicted density, and the lower-right panel, these differences
as a histogram that includes all of the grid points.  The hydrostatic
model is generally more concentrated towards the equatorial plane than
the fixed model, and this is particularly clear from the histogram
of differences. This central concentration is a result of the cool,
equatorial region that develops in a such a high density disk (see
next paragraph).

\begin{figure}
\epsscale{1.0}
\plotone{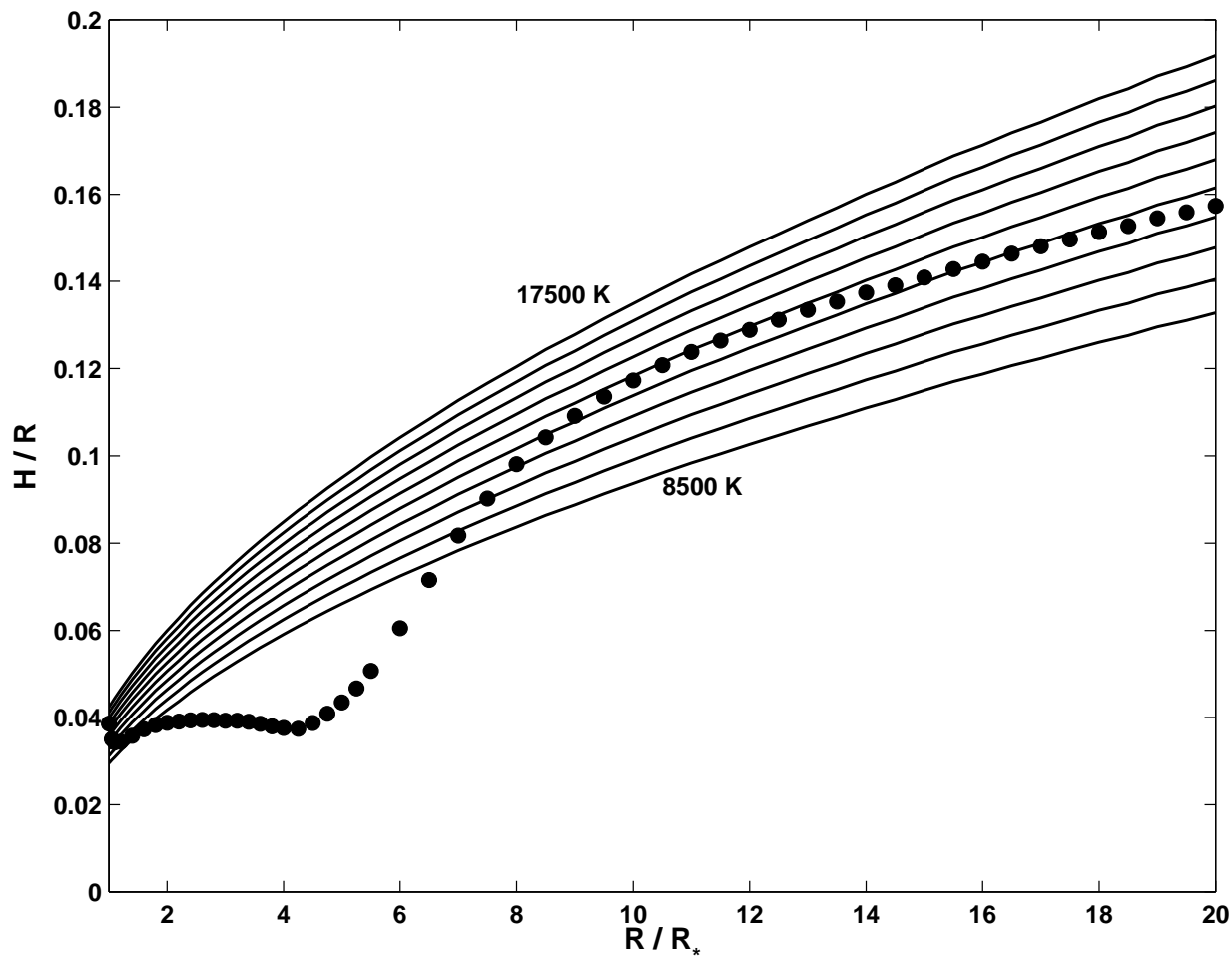}
\caption{
Vertical disk scale height as a function of radial distance for
a B0 model with $\rho_o=5\,10^{-11}\,\rm g\,cm^{-3}$ and $n=3.5$.
The solid lines are scale heights for fixed models (Eqns~\ref{eq:rho}
and \ref{eq:scale_height}) with isothermal temperatures ranging from 8500
to $17500\;$K in steps of $1000\;$K (moving upward in the figure). The
filled circles are the scale heights of a hydrostatic model of the same
$\rho_o$ and $n$ found by locating the vertical height at which the
density dropped to $1/e$ of its value in the equatorial plane.
\label{fig:hvsr}}
\vspace{0.1in}
\end{figure}

%
%
Figure~\ref{fig:hvsr} compares the predicted vertical disk scale heights
between the fixed and hydrostatic disks. Scale heights for fixed disks
follow from Eq.~\ref{eq:scale_height} and are shown for isothermal
temperatures ranging from $8500$ to $17500\;$K. Scale heights for the
hydrostatic disk are found numerically by locating the point above the
disk at which the density falls to $1/e$ of its value in the equatorial
plane. The cool equatorial region that develops close to the star (see
additional discussion below) is clearly reflected in the hydrostatic disk
scale height for $R/R_* \leq 5$. Beyond this, the scale height rapidly
rises, and by $R/R_* \approx 10$, it closely matches the fixed, isothermal model
corresponding to $T_o = 13,500\;$K. Thus in the region close
to the star, $R/R_*\leq 10$, the scale heights and their variation with
$R$ are not well represented by any of the isothermal models.

One additional subtle point is that the hydrostatic disk is somewhat less
massive than the fixed disk.  For the fixed model, the mass follows
directly from Eq.~\ref{eq:rho} and the adopted parameters.  However,
for the hydrostatic disk, the mass follows the density adopted in
the equatorial plane (Eq.~\ref{eq:PL}) and the hydrostatic equilibrium
solution. In this case, the mass of the fixed disk is $2.78\cdot10^{-8}$
solar masses whereas the mass of the hydrostatic disk is about 20\%
less at $2.16\cdot10^{-8}$ solar masses.

\begin{figure}
\epsscale{1.0}
\plotone{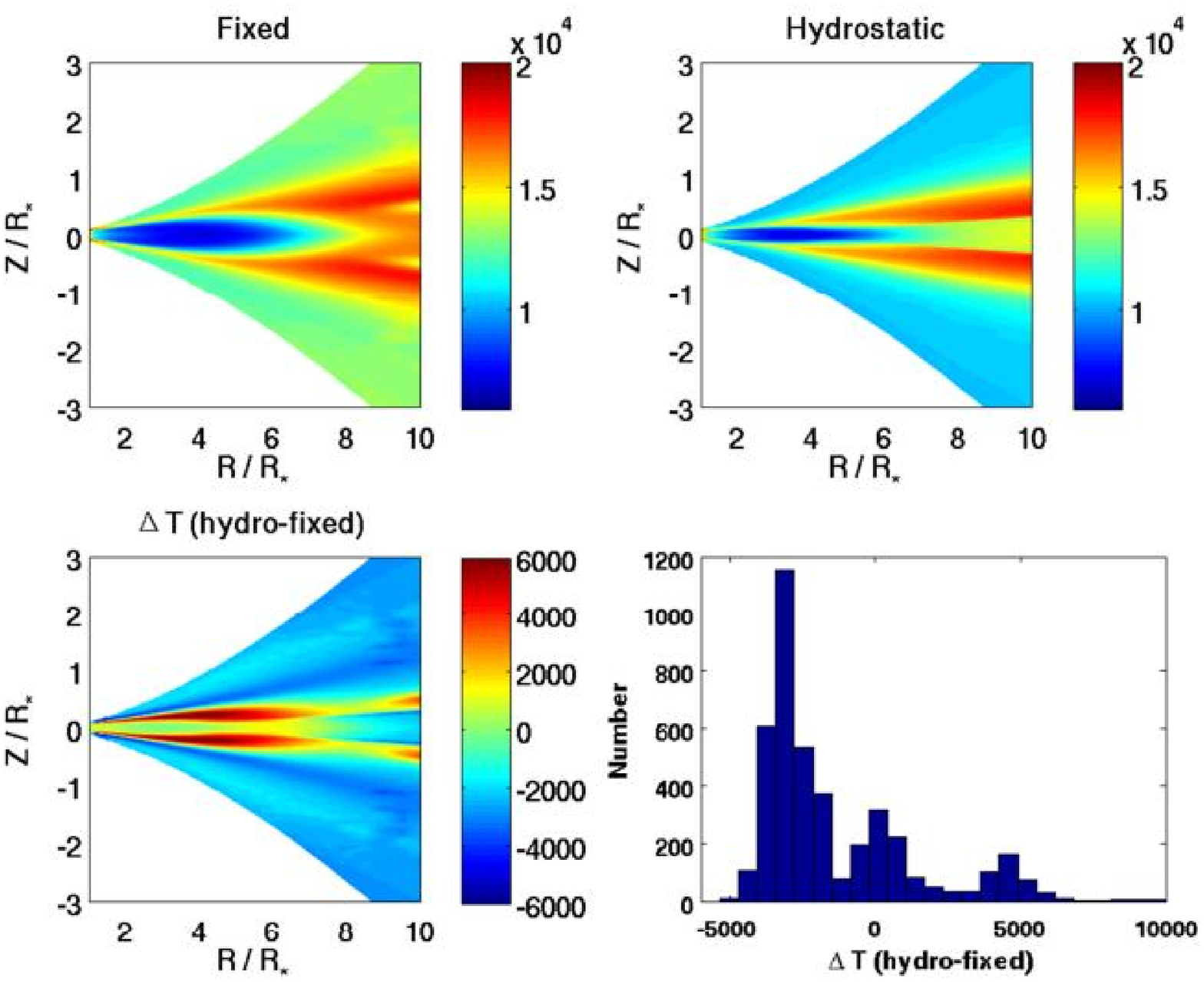}
\caption{The temperature structure of a fixed (upper
left) and hydrostatic (upper right) disk with $\rho_o=5\cdot10^{-11}\,
\rm g\,cm^{-3}$ and $n=3.5$ for the B0 model.
The lower-left panel shows the temperature differences directly,
while the lower-right panel gives a histogram
of these differences over all 4200 grid points.\label{fig:egtemp}}
\vspace{0.1in}
\end{figure}

Figure~\ref{fig:egtemp} compares the predicted temperature
structure of both disks. As shown in the lower panels, there are
significant differences in temperature. Due to the large initial
density ($\rho_o=5\cdot10^{-11}\;\rm g\,cm^{-3}$), both the fixed and
hydrostatic models have a cool, equatorial zone close to the star. Above
and below this cool zone are hotter sheaths which can still be directly
illuminated by at least part of the central star.  Large temperature
differences between the fixed and hydrostatic models can result from the
differing locations of these hot sheathes as the hydrostatic model is
more concentrated in density towards the equatorial plane. There is also a
significant temperature difference between the two models at the location
of the optically thin gas far above the equatorial plane. This is a result
of the very large density difference between the fixed and hydrostatic
models in this region as illustrated in Figure~\ref{fig:egden}.

\begin{figure}
\epsscale{1.0}
\plotone{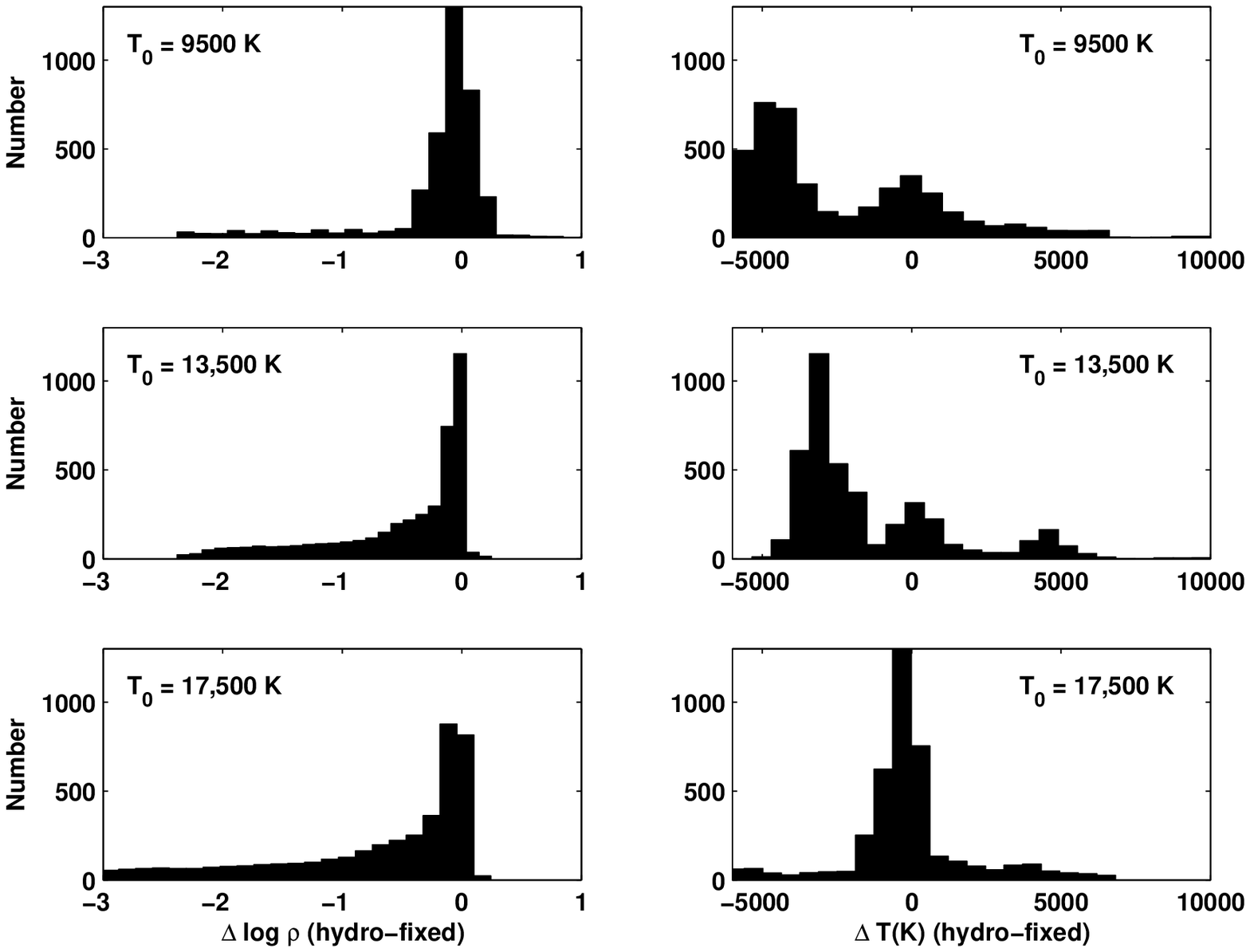}
\caption{Histograms of the density (left panels) and temperature (right panels) differences
between fixed and hydrostatic models for $\rho_o=5\cdot10^{-11}\,\rm g\,cm^{-3}$ and $n=3.5$
for the B0 model.
The fixed models assume $T_o=9,500\;$K (top row), $13,500\;$K (middle row), and
$17,500\;$K (bottom row). \label{fig:histB0}}
\vspace{0.1in}
\end{figure}

In this comparison, one might question the choice of $T_o=13,500\;$K
adopted for the fixed model. To address this point, an additional
set of fixed calculations was performed by varying $T_o$ from
$9,500$ to $17,500\;$K with all of the other model parameters held
fixed. Figure~\ref{fig:histB0} summarizes these results by giving
histograms of the differences in temperature and in the logarithm of the
density over the entire grid. As can be seen from Figure~\ref{fig:histB0},
the density distribution is best represented by the coolest model,
$T_o=9,500\;$K. This is not surprising as $\rho_o$ is large enough that
a cool equatorial region develops with temperatures as low as $8,000\;$K
in some regions. Conversely, Figure~\ref{fig:histB0} shows that the
temperature is best represented by the hottest model, $T_o=17,500\;$K.
This result, however, is a bit deceptive. As will be shown in the next
section, it is the coolest model which does the best job in reproducing
the observables. The large temperature differences in the $T_o=9,500\;$K
model often result from the misalignment in the height of the cool gas
and hot sheaths.

\begin{figure}
\epsscale{1.0}
\plotone{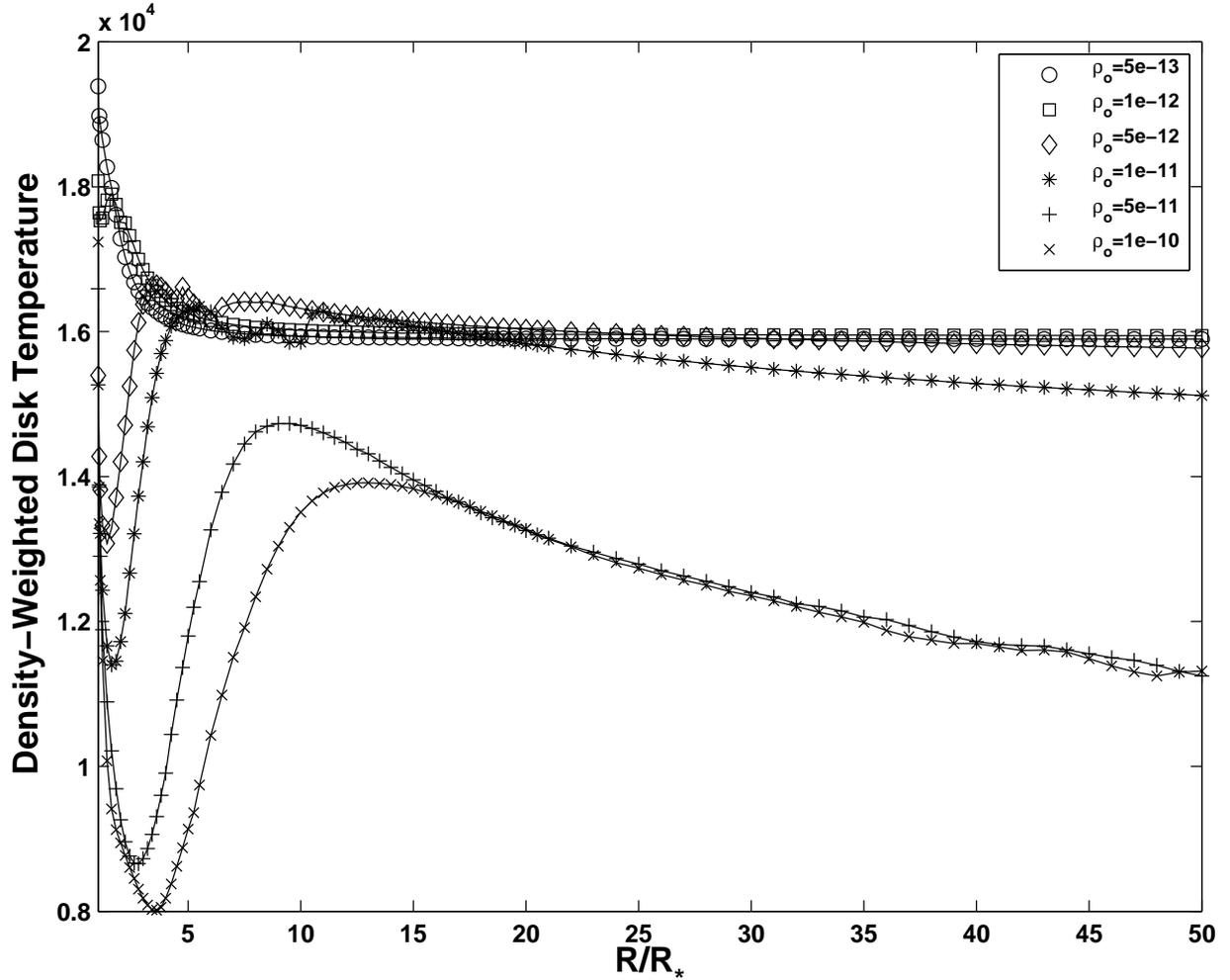}
\caption{The density-averaged temperature as a function of radial distance
for six hydrostatic models corresponding to values of $\rho_o$ ranging
from $5.0\cdot10^{-13}\;\rm g\,cm^{-3}$ to $1.0\cdot10^{-10}\;\rm g\,cm^{-3}$
(see legend) for the B0 model.\label{fig:tav}}
\vspace{0.1in}
\end{figure}

\begin{figure}
\epsscale{1.0}
\plotone{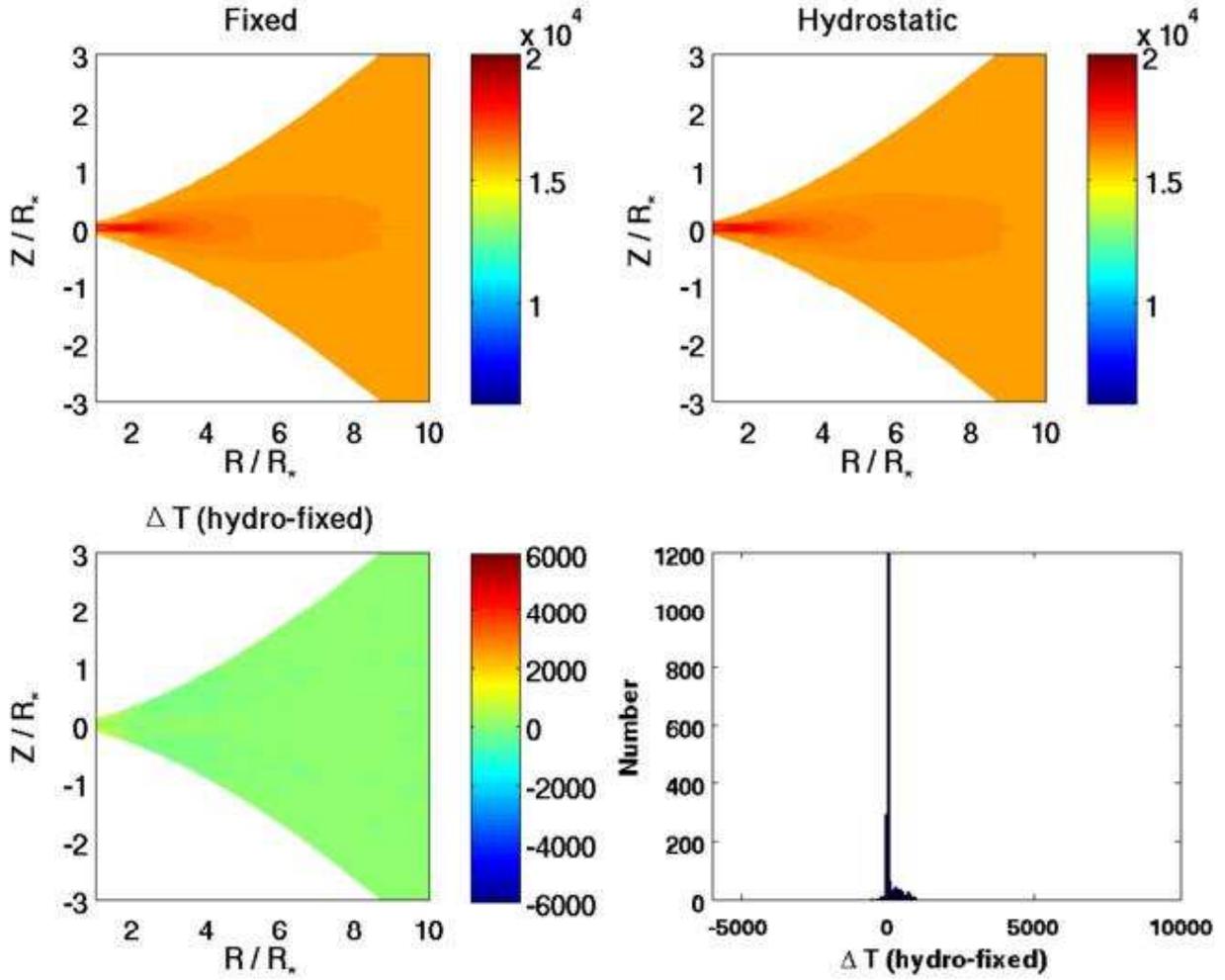}
\caption{Comparison of the disk temperatures between a fixed (upper
left) and hydrostatic (upper right) model with for $\rho_o=1\cdot10^{-12}\,
\rm g\,cm^{-3}$ and $n=3.5$ for the B0 model.
The fixed model assumes $T_o=13,500\;$K.
\label{fig:egtemp_lowp}}
\vspace{0.1in}
\end{figure}

\begin{figure}
\epsscale{1.0}
\plotone{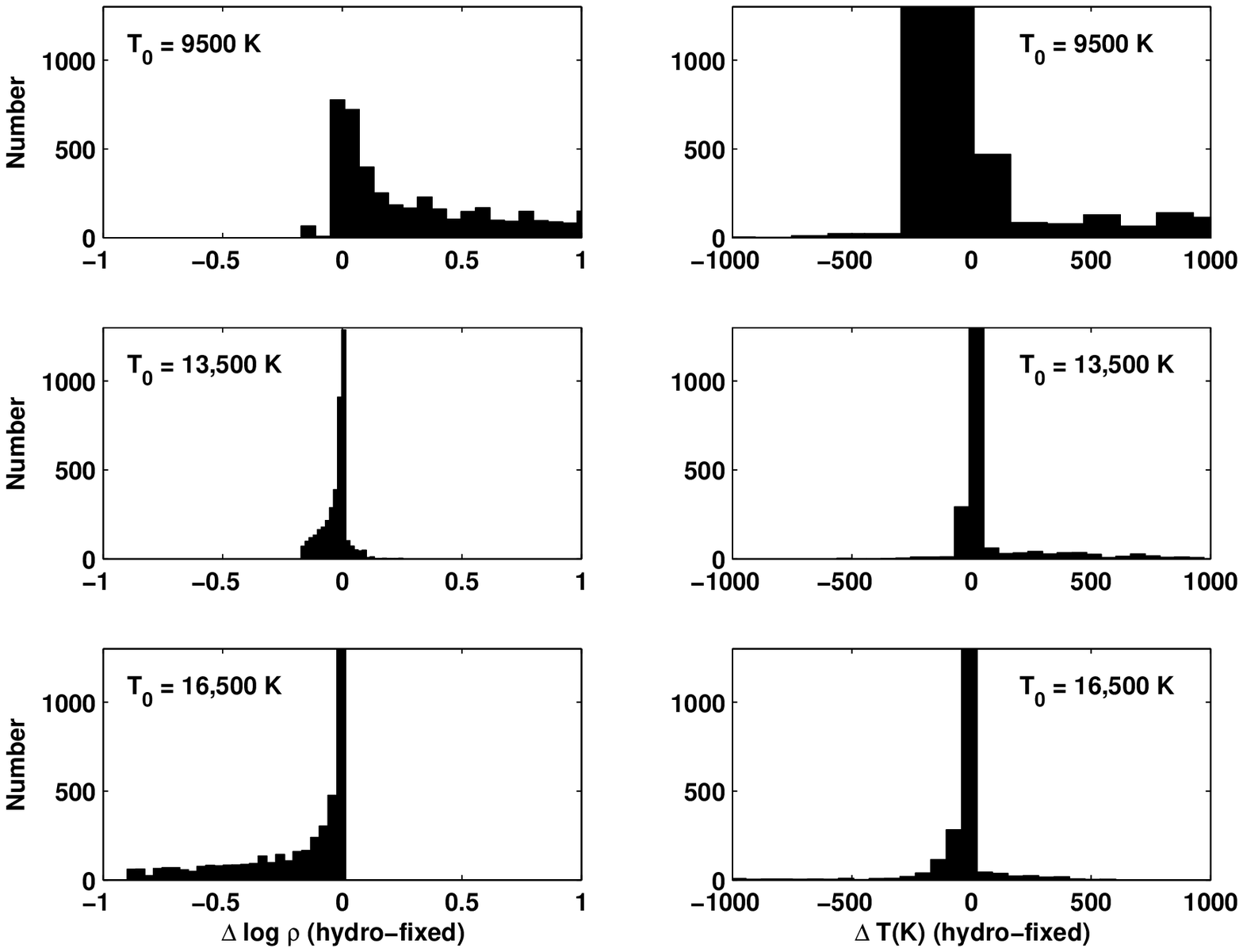}
\caption{Histograms of the density (left panels) and temperature (right panels) differences
between fixed and hydrostatic models for $\rho_o=1\cdot10^{-12}\,\rm g\,cm^{-3}$ 
and $n=3.5$ for the B0 model.
The fixed models assume $T_o=9,500\;$K (top row), $13,500\;$K (middle row), and
$17,500\;$K (bottom row). \label{fig:histB0_lowp}}
\vspace{0.1in}
\end{figure}

The previous large differences between the fixed and hydrostatic
models reflect the poor job the assumption of an isothermal $T_o$
does in representing the density structure of a disk that develops a cool equatorial
region. Figure~\ref{fig:tav} illustrates the development of this
cool region, for the B0 spectral type, as the overall disk density is increased. Plotted is the
density-weighted temperature,
\begin{equation}
\label{eq:trho}
<T_{\rm Disk}(R)> \equiv \frac{\int_{0}^{Z_{\rm max}} \rho(R,Z)\,T(R,Z) \,dZ}
  {\int_{0}^{Z_{\rm max}} \rho(R,Z)\,dZ} \;,
\end{equation}
as a function of radial distance for six hydrostatic models with
initial densities, $\rho_o$, ranging from $5.0\cdot10^{-13}$ to
$1.0\cdot10^{-10}\;\rm g\,cm^{-3}$.  All models assumed $n=3.5$.  This
figure illustrates that models with $\rho_o\lesssim 5.0\cdot10^{-12}\;\rm
g\,cm^{-3}$, do not have an extensive cool, equatorial zone near
the star where $T(R,Z)\lesssim 10,000\;$K whereas denser models
rapidly develop such a region.  This lack of a strong temperature
gradient for the less dense models suggests that the isothermal
approximation of Eq.~\ref{eq:rho} is adequate to represent their
density structure. This is borne out by Figure~\ref{fig:egtemp_lowp}
which compares the disk temperatures of a low-density
($\rho_o=1.0\cdot10^{-12}\;\rm g\,cm^{-3}$) hydrostatic model with
a fixed model computed with the same $\rho_o$ and $T_o=13,500\;$K.
There is little temperature difference between the
two models.  Figure~\ref{fig:histB0_lowp} compares (as histograms)
the predicted disk temperatures and densities for three choices of the
$T_o$ parameter, $9,500$, $13,500$ and $17,500\;$K, with the hydrostatic
model. As is clear from this figure, $T_o=13,500\;$K does an adequate
job of reproducing both the densities and temperatures in the disk.

The general conclusion is that the isothermal, fixed $T_o$ models are
appropriate for low-density disks provided a reasonable choice for $T_o$
is made.  However, denser disks develop a cool equatorial region and
the disk densities and the resultant temperatures cannot be computed
assuming a fixed, isothermal density structure. This result is confirmed
by similar calculations for the disks surrounding the two later spectral
types of Table~\ref{tab:stellar_param}. However, rather than present
similar plots and histograms for these models, we will proceed to the next
section where a systematic comparison of the predicted density-averaged
disk temperatures is made between fixed and hydrostatic models.

%

\subsection{Density-Averaged Disk Temperatures}

To further investigate the reliability of the fixed density models,
we examine how well they can predict the global, density-weighted, average disk
temperature, defined as
\begin{equation}
\label{eq:tav_defn}
<T_{\rm Disk}> \equiv \frac{1}{M_{\rm Disk}} \int_{V} \rho\,T\,dV \;.
\end{equation}
Here $M_{\rm Disk}$ is the total mass of the disk, and the integral is
over the total volume of the disk.

\begin{figure}
\epsscale{1.0}
\plotone{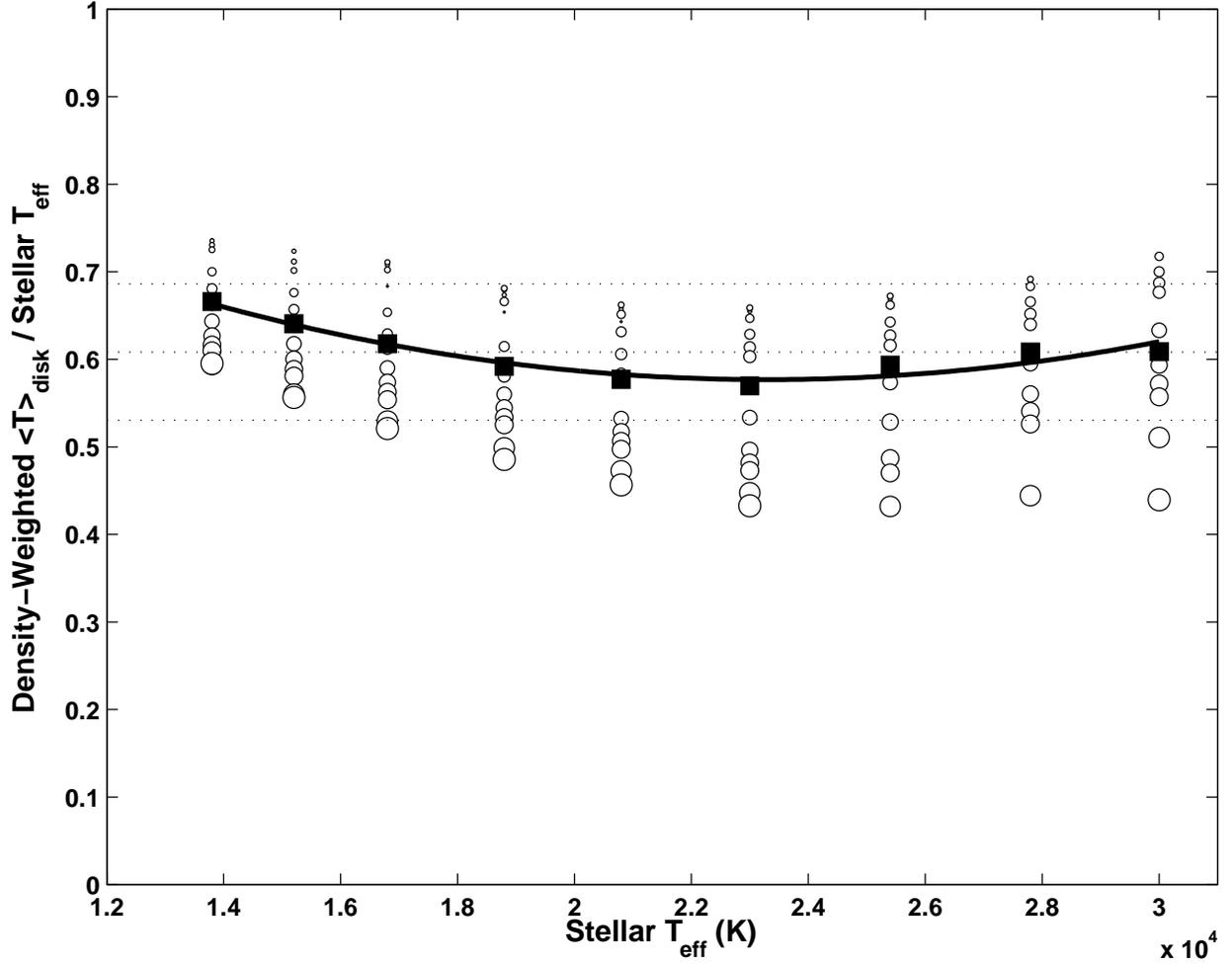}
\caption{Density-weighted disk temperature (Equation~\ref{eq:tav_defn}) as a 
function of the central star's $T_{\rm eff}$. At each $T_{\rm eff}$, the open
circles represent disks with different base densities, $\rho_o$, with the 
size of the symbol signifying the size of $\rho_o$. The smallest circles correspond
to $\rho_o=2.5\cdot10^{-13}\;\rm g\,cm^{-3}$ and the largest circles,
to $\rho_o=5.0\cdot10^{-10}\;\rm g\,cm^{-3}$. The filled squares are the
(unweighted) average temperatures, and the solid line is the quadratic fit
given by Equation~(\ref{eq:tav_fit}). \label{fig:trho_hydro}}
\vspace{0.1in}
\end{figure}

Equation~\ref{eq:tav_defn} was computed for 9 central stars, ranging
in $T_{\rm eff}$ from $13,800\;$K to $30,000\;$K in order to cover the
full range of Be star spectral types. For each central star,
14 disk models were computed for different choices of the base disk
density\footnote{The
densities used were $\rho_o=2.5$, $5.0$, $7.5 \cdot10^{-13}\;\rm g\,cm^{-3}$;
$1.0$, $2.5$, $5.0$, $7.5\cdot10^{-12}\;\rm g\,cm^{-3}$;
$1.0$, $2.5$, $5.0$, $7.5\cdot10^{-11}\;\rm g\,cm^{-3}$; and
$1.0$, $2.5$, $5.0 \cdot10^{-10}\;\rm g\,cm^{-3}$.},
$\rho_o$, covering the range $\rho_o=2.5\cdot10^{-13}\;\rm
g\,cm^{-3}$ to $5.0\cdot10^{-10}\;\rm g\,cm^{-3}$.
All models assumed an
$n=3.5$ radial power-law index. In total, 126 hydrostatic disk models
were computed. Figure~\ref{fig:trho_hydro} shows the results with the
density-weighted, disk temperature expressed as a fraction of the stellar
$T_{\rm eff}$. The (unweighted) average ratio at each $T_{\rm eff}$ (shown
in the figure by the filled squares) is well-fit by the quadratic relation
\begin{equation} 
\frac{<T_{\rm Disk}>}{T_{\rm eff}} = 0.096\,T^2_{4}
- 0.448\,T_4 + 1.098 \;, \label{eq:tav_fit}
\end{equation} 
where $T_4$ is the stellar effective temperature in units of $10^4\;$K.
Over the $T_{\rm eff}$ range of the Be stars, the average ratio ranges
between 0.55 and 0.65, suggesting some validity to the often-used
approximation that a good estimate for an (isothermal) disk temperature
is a constant fraction of the stellar $T_{\rm eff}$ (for example,
\cite{wat87} adopted $<T_{\rm Disk}>=0.8\,T_{\rm eff}$). Similar ratios are
found throughout the literature, including some based on sophisticated
modeling, such as that of \cite{car06} who found a ratio of 0.6 for
moderate density disks surrounding a B3~IV star. The current work finds
that $<T_{\rm Disk}> / T_{\rm eff}\approx 0.6$ is a reasonable fit to the
average trend over the entire range of Be stars, with the quadratic fit of
Eqn~(\ref{eq:tav_fit}) representing a marginal improvement. However,
the current work also makes clear that at each individual $T_{\rm eff}$,
the scatter about this average is large and depends in a systematic way
on the disk density $\rho_o$. Increasing the overall density drives the
density-weighted temperatures to lower values due to the development
of the cool equatorial zone noted in Figure~\ref{fig:tav}. As shown in
Figure~\ref{fig:trho_hydro}, ratios as low as 0.45 can be predicted for
dense disks, and ratios as high as 0.7 for rarefied disks.

\begin{figure}
\epsscale{1.0}
\plotone{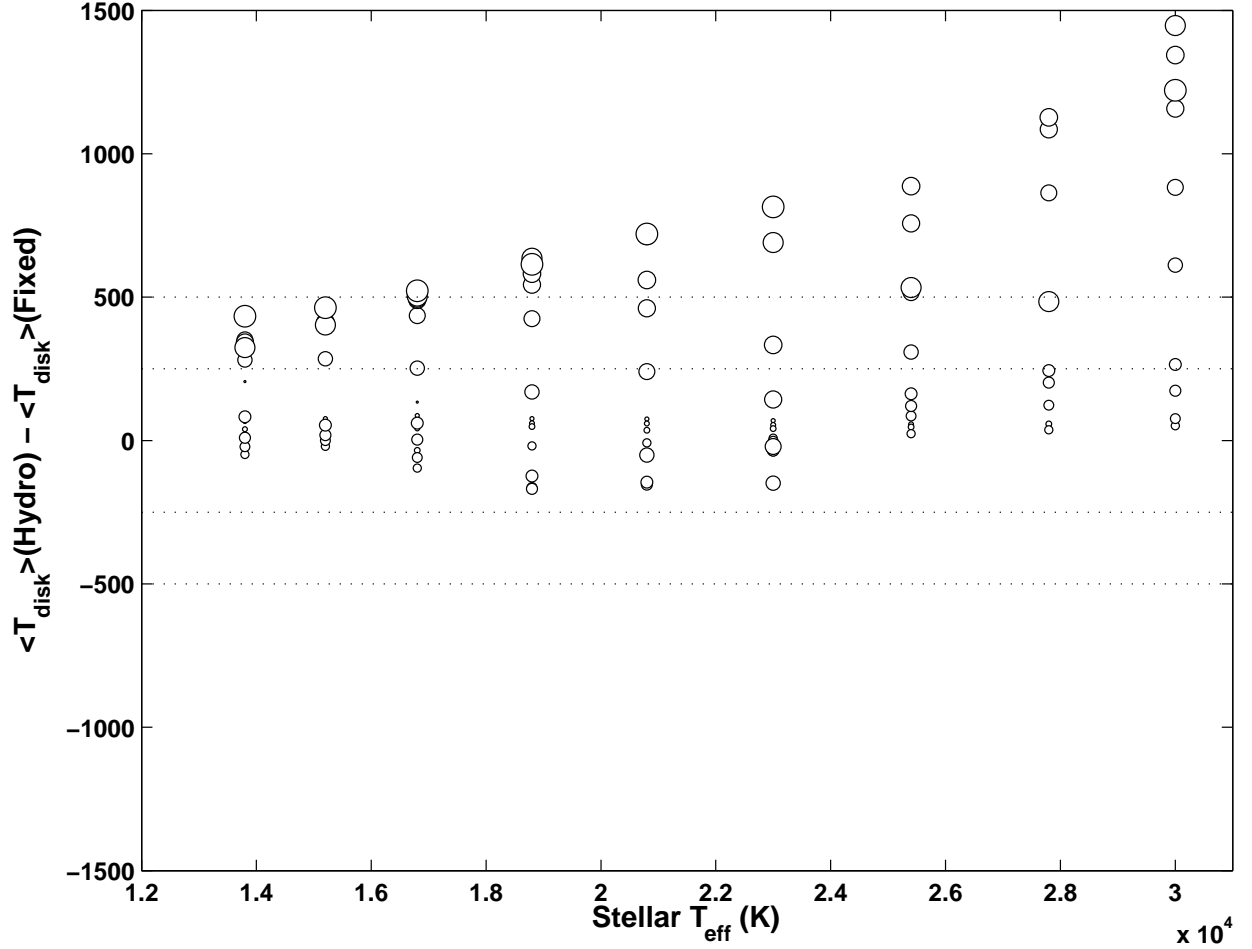}
\caption{Difference between the density-weighted disk temperature predicted
by the fixed and hydrostatic models. For the fixed models, the $T_o$
parameter was chosen to be the hydrostatic prediction for $<T_{\rm Disk}>$.
As in Figure~(\ref{fig:trho_hydro}), the open symbols represent the different
base densities chosen for each disk.\label{fig:trho_comp}}
\vspace{0.1in}
\end{figure}

To examine the accuracy of the fixed models in predicting the
density-weighted disk temperature over a wide range of central effective
temperatures and disk densities, the 126 models were rerun but
this time as fixed density models with $T_o$ (see Equation~\ref{eq:alpha}
and discussion) chosen to be equal to the $<T_{\rm Disk}>$ predicted
by the corresponding hydrostatic model.  Figure~\ref{fig:trho_comp}
compares the $<T_{\rm Disk}>$ predicted by the fixed models to the
hydrostatic results. As can be seen from the figure, the fixed models
with an appropriately chosen $T_o$ can reproduce the $<T_{\rm Disk}>$ of
the hydrostatic models to within $\pm500\;$K over a wide range of $T_{\rm
eff}$ and $\rho_o$.  This is particularly true for the models with low
values for $\rho_o$, which scatter well within $\pm250\;$K. However,
larger deviations occur for the denser disks and, for $T_{\rm eff} >
18,000\;$K, the fixed models for the largest densities all predict
hotter $<T_{\rm Disk}>$ values by up to $1500\;$K.

Despite this reasonable agreement, it is very important to next see if fixed
models can be used to successfully predict more readily observed
diagnostics such as H$\alpha$ line profiles and infrared excesses.

\subsection{Observational Diagnostics}

To demonstrate more direct observational consequences of consistent
vertical hydrostatic equilibrium, we considered a disk with a density
of $\rho_o=5.0\cdot10^{-11}\;\rm g\,cm^{-3}$ and $n=3.5$. As noted
in Section~\ref{sec:tandrho}, this density is large enough that a cool
equatorial region forms close to the star where the vertical temperature
distribution is far from isothermal.  Radiative equilibrium solutions
were found for the fixed density structure of Eq.~\ref{eq:rho} for
several values of the $T_o$ parameter. Observable quantities were
computed for each of these fixed models to see if any could reproduce the
observational predictions of the hydrostatic disk with the same $\rho_o$
and $n$. This process was repeated for each of the three stars given in
Table~\ref{tab:stellar_param} which correspond spectral types B0, B2,
and B5.

%
%
The first observational diagnostic considered is the predicted
H$\alpha$ line profile corresponding to each model.  The H$\alpha$
line profile was obtained by solving the transfer equation using the
hydrogen level populations predicted by {\sc bedisk}. In all cases, a
viewing angle of $i=35\,$ degrees was chosen (where $i=0$ corresponds
to a pole-on star) and the equatorial velocity was set to $325\;\rm
km\,s^{-1}$. This results in a projected stellar rotation velocity of
$v\,\sin\,i=186\;\rm km\,s^{-1}$. An equatorial velocity of $325\;\rm
km\,s^{-1}$ is approximately 70\% of the critical rotation velocities
given the parameters of Table~\ref{tab:stellar_param}.  Pure Keplerian
rotation, with zero outflow velocity, was assumed for the disk.

To represent the stellar disk, an LTE, photospheric, H$\alpha$ line
profile was used corresponding to the $T_{\rm eff}$ and $\log(g)$
of Table~\ref{tab:stellar_param}, The profile for each element of the
stellar surface was shifted by its projected radial velocity, resulting
in a rotationally broadened photospheric profile.  For the calculation
of the H$\alpha$ emissivity and opacity in the disk, the Stark profiles
of \citet{bar03} were used.

\begin{figure}
\epsscale{1.0}
\plotone{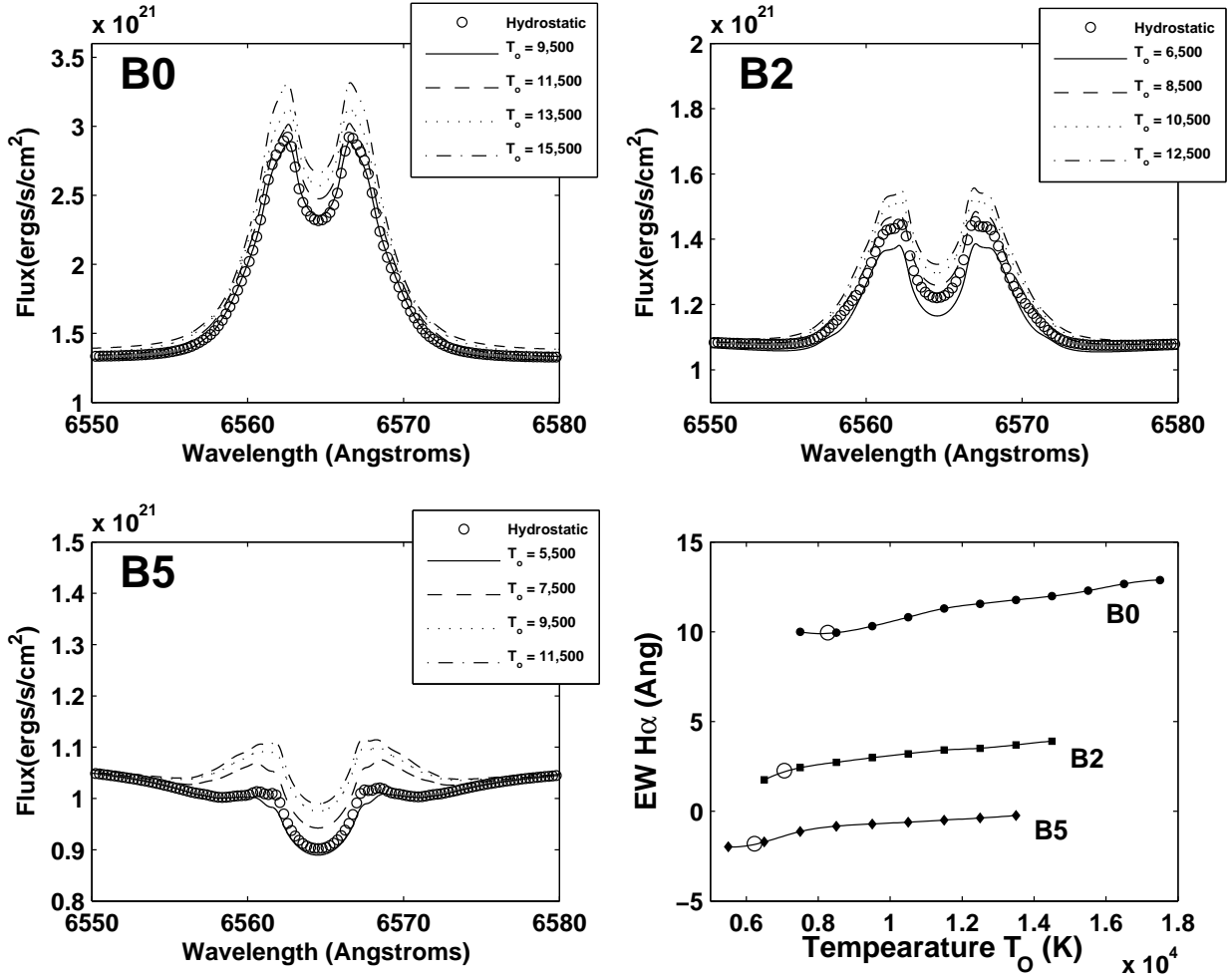}
\caption{Predicted H$\alpha$ line profiles for hydrostatic models and 
fixed models (assuming various $T_o$) for three stellar spectral types 
(upper panels and lower left panel). The 
variation of the H$\alpha$ equivalent width (in \AA) for these stars as a function of
$T_o$ for the fixed models is also shown in the lower right panel. The
equivalent width predicted by the hydrostatic model associated with these 
stars is shown as an open circle.
\label{fig:ha}}
\vspace{0.1in}
\end{figure}

Figure~\ref{fig:ha} shows the resultant H$\alpha$ line profiles
predicted by the hydrostatic and fixed models for each of the three
spectral types considered.  The fourth panel shows the variation with
$T_o$ of the total H$\alpha$ equivalent width of the fixed models.
The equivalent widths of the consistent hydrostatic models are also shown.
The H$\alpha$ equivalent width increases with $T_o$ for the fixed models
and the increase is $\approx\,30$\% for a factor of two increase in
$T_o$. Typically, the fixed model with (nearly) the lowest value of $T_o$
is most successful in matching the hydrostatic prediction. This again
reflects the influence of the cool equatorial region that develops in
the higher-density disks. The H$\alpha$ emissivity is controlled mainly
by the high temperature ``sheaths" above and below the equatorial plane
in the inner disk. The lower $T_o$ values better represent the actual
inner disk scale heights thus placing the hot sheaths closer the location
predicted by the hydrostatic models (see Figure~\ref{fig:egtemp}).

\begin{figure}
\epsscale{1.0}
\plotone{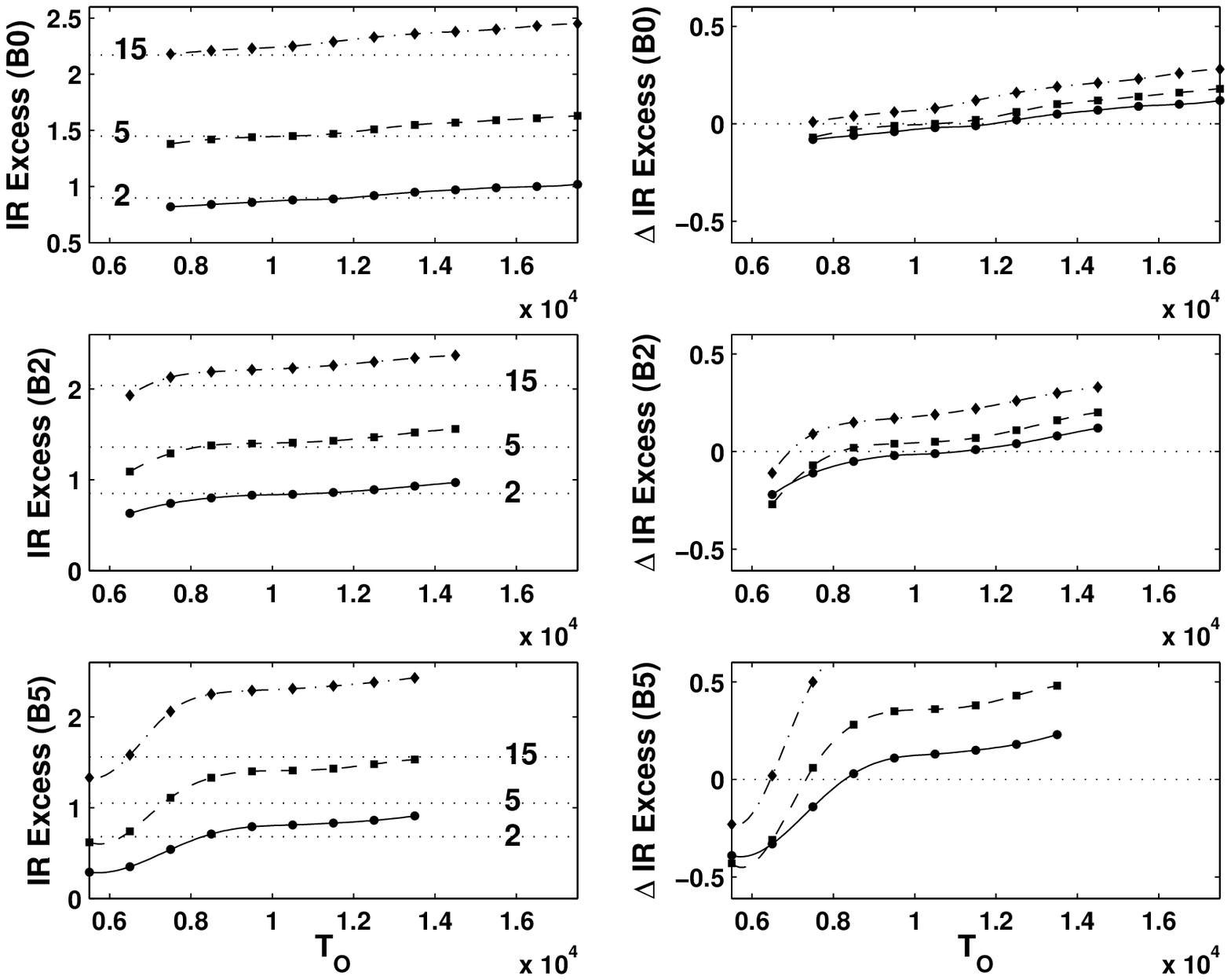}
\caption{Predicted IR
excess for a series of fixed disk models computed by varying $T_o$ for
three stellar spectral types. Absolute IR excesses are shown on the panels
on the left. The line styles and symbols correspond to wavelengths of 2$\,\mu$m (solid
line, circles), 5$\,\mu$m (dashed line, squares), and 15$\,\mu$m (dash-dot
line, diamonds). The predictions of the hydrostatic models are shown as a dotted
line with the wavelength in $\mu$m as indicated. The differences in the IR excess,
fixed minus hydrostatic, are shown in the right-hand panel. The line styles have
the same interpretation except that the dotted line in each shows the zero
magnitude difference for reference. \label{fig:ir}}
\vspace{0.1in}
\end{figure}

Another observational diagnostic is the infrared excess predicted
by the models. Infrared excesses (expressed in magnitudes),
relative to the underlying photospheric contribution, are shown
in Figure~\ref{fig:ir}. Again, the excess is found by solving the
transfer equation along a series of rays threading the disk system. To
represent the star, an LTE, line-blanketed stellar atmosphere was used
corresponding to the parameters listed in Table~\ref{tab:stellar_param}.
Solutions were obtained at three wavelengths in the near-infrared,
2, 5, and 15$\;\mu$m. The predicted IR excess for a series of fixed
disk models with varying $T_o$ are shown in Figure~\ref{fig:ir}.
Also shown in the figure are the predictions of the hydrostatic models
with the same $\rho_o$ and $n$ for each spectral type. For the earlier
spectral types, the differences are generally small, within a few
tenths of a magnitude, for plausible choices of $T_o$.  However, larger
differences are predicted for the latest spectral type considered, B5.
Close inspection of this figure also shows that there is, in general,
no single choice for $T_o$ that will reproduce the IR excess at the
three wavelengths considered. Typically a higher $T_o$ is required to
match the excess at a shorter wavelength; this effect is particularly
clear in the B5 model where the magnitude differences are largest.

\section{Conclusions}

In this work, we have compared predicted disk temperatures, densities,
H$\alpha$ line profiles and equivalent widths, and near-IR excesses
between a set of Be disk models computed in consistent radiative and
vertical hydrostatic equilibrium and a set of corresponding radiative
equilibrium models that assumed a fixed density structure. Large
differences between the predicted temperatures and densities can occur
between the hydrostatic and fixed models when the density is large enough
that the disk develops a cool equatorial region close to the star. In this
case, there seems to be no choice for the single, isothermal temperature
$T_o$ that characterizes the fixed density structure that will yield a
model matching the observational signatures of a consistent model.

These conclusions are particularly important given the way such sets
of Be disk models are used to compare to observations and extract disk
parameters. Typically, a grid of models is computed for a wide range
of $\rho_o$ with a fixed density structure parametrized by a single
temperature $T_o$ \citep[see for example][]{sig07}.  This grid is then
used to compared to observations of, say, H$\alpha$ line profiles or
IR excesses to select the most appropriate disk parameters. As shown in
this work, such a grid will contain a systematic error in the predictions
for high $\rho_o$ where a cool equatorial region in the disk develops;
such models are poorly approximated by disks with a density structure
fixed a priori.  To extract the ``true" distribution of disk parameters,
within the context of a thin disk in vertical hydrostatic equilibrium,
a grid which consistently enforces both radiative and vertical hydrostatic
equilibrium should be employed.

\acknowledgments
We would like to thank Chris Tycner and Jon Bjorkman for many helpful
discussions.  This work is supported by the Canadian Natural Sciences
and Engineering Research Council through Discovery Grants to TAAS and CEJ.

\end{document}